\documentclass[12pt]{article}

\usepackage{graphicx}

\usepackage{amssymb}
\usepackage{amsmath}
\usepackage{amscd}
\usepackage{latexsym}

\newcommand{\be}{\begin{equation}}
\newcommand{\ee}{\end{equation}}
\newcommand{\Dlt}{\Delta}
\newcommand{\dlt}{\delta}
\newcommand{\prt}{\partial}
\newcommand{\br}{{\bf r}}
\newcommand{\bk}{{\bf k}}
\newcommand{\bE}{{\bf E}}
\newcommand{\bH}{{\bf H}}
\newcommand{\bP}{{\bf P}}
\newcommand{\bA}{{\bf A}}
\newcommand{\bS}{{\bf S}}
\newcommand{\bd}{{\bf d}}
\newcommand{\bJ}{{\bf J}}
\newcommand{\bj}{{\bf j}}
\newcommand{\bt}{\beta}

\newcommand{\ep}{\varepsilon}
\newcommand{\al}{\alpha}
\newcommand{\ra}{\rightarrow}

\newcommand{\gm}{\gamma}
\newcommand{\om}{\omega}
\newcommand{\Om}{\Omega}
\newcommand{\Gm}{\Gamma}

\newcommand{\lbd}{\lambda}

\newcommand{\lgl}{\langle}
\newcommand{\rgl}{\rangle}

\begin{document}

\begin{center}
{\Large{\bf Dynamics of quantum-dot superradiance} \\ [10mm]

V.I. Yukalov$^1$ and E.P. Yukalova$^2$} \\ [5mm]

{\it $^1$Bogolubov Laboratory of Theoretical Physics, \\
  Joint Institute for Nuclear Research, Dubna 141980, Russia} \\ [5mm]

{\it $^2$Laboratory of Information Technologies \\
  Joint Institute for Nuclear Research, Dubna 141980, Russia}

\end{center}

\vskip 3cm

\begin{abstract}

The possibility of realizing the superradiant regime of
electromagnetic emission by the assembly of quantum dots is
considered. The overall dynamical process is analyzed in detail.
It is shown that there can occur several qualitatively
different stages of evolution. The process starts with dipolar
waves triggering the spontaneous radiation of individual dots.
This corresponds to the {\it fluctuation stage}, when the dots
are not yet noticeably correlated with each other. The second is
the {\it quantum stage}, when the dot interactions through the
common radiation field become more important, but the coherence
is not yet developed. The third is the {\it coherent stage},
when the dots radiate coherently, emitting a superradiant pulse.
After the superradiant pulse, the system of dots relaxes to an
incoherent state in the {\it relaxation stage}. If there is no 
external permanent pumping, or the effective dot interactions are
weak, the system tends to a stationary state during the last
{\it stationary stage}, when coherence dies out to a low,
practically negligible, level. In the case of permanent pumping,
there exists the sixth stage of {\it pulsing superradiance}, when
the system of dots emits separate coherent pulses.

\end{abstract}

\vskip 2cm


\newpage

\section{Introduction}

Superradiance is the effect of self-organized collective coherent
radiation by an ensemble of radiators. The phenomenon of optical
superradiance is well known for atoms and molecules, being
described in numerous publications (see., e.g., books [1-3]).
Optical superradiance from Bose-Einstein condensed atoms has also
been investigated [4]. There also exists spin superradiance
produced by spin systems, such as nuclei [5-8] or magnetic
molecules [9-12].  It has been suggested [13] that the assemblies
of quantum dots or wells could also be arranged so that to produce
superradiance. This can become possible when the distance between
the neighboring quantum nanostructures is smaller than the
radiation wavelength. In that case, there appears an effective
interaction between the radiators, due to common radiation field.
The modern technology of preparing materials with quantum dots
allows for the fabrication of the quantum dot assemblies with the
density of quantum dots sufficient for the appearance of such an
effective interaction [13-17], which has been detected
experimentally [18].

Quantum dots, whose electrons or holes are confined in all three
spatial dimensions, have many properties [19-22] that make them
similar to atoms, because of which quantum dots are often called
"artificial atoms". One of the main such properties is the
existence of discrete energy levels, whose shell structure can
be adjusted by design. Some of these properties are shared by
quantum wells confining electrons or holes in one dimension and
allowing free propagation in two dimensions. For concreteness,
we shall concentrate in what follows on the consideration of
quantum dots.

Transferring electrons from the ground-state energy level to
an excited level creates a hole. The interacting pair of an
excited electron and a hole forms an exciton. The electron-hole
recombination is accompanied by the radiation of electromagnetic
field, which is in a very close analogy with the radiation of
excited atoms. This is why it was reasonable to assume [23]
that an ensemble of quantum dots can be employed for the creation
of quantum-dot lasers. Several types of quantum-dot lasers have
been demonstrated since then, based on semiconductor
heterojunctions [24-29] and photonic crystals [30-33].
Quantum-well lasers have also been realized. But the latter, 
because of thermal occupation of the the quasi-2D continuum, are
essentially more sensitive to temperature changes, which makes 
their emission less monochromatic than that of quantum dots. 
Quantum-dot lasers also enjoy a lower current-density lasing 
threshold. Because of these differences, quantum dots look more 
suitable for the use as radiating devices.

When quantum dots are fabricated in the process of epitaxial
growth, they are usually characterized by a noticeable size
dispersion, which results in the related inhomogeneous
broadening. The latter may hinder the possibility of achieving
efficient dot interactions through the common radiation field and,
as a result, destroying  the cooperative character of emission,
thus, suppressing coherence [15,34,35]. However the impressive
recent progress in controlling quantum-dot parameters [18,27] 
makes it now feasible to create quantum dots so that they allow
for the appearance of effective dot interactions through 
electromagnetic field, which is a prerequisite for achieving 
collective coherent radiation.

Because of the technological feasibility of fabricating
semiconductor samples with sufficiently dense and uniform
quantum dots, it should be possible to realize the conditions
for their superradiant emission. It is the aim of the present
paper to consider the quantum-dot superradiance. The main goal
is to develop an accurate and detailed description of all
stages of the superradiant dynamics, starting from the initial
stage, when collective effects are yet weak, to the coherent
stage of radiation, and further to the end of the whole process.
The developed theory makes it possible to describe different
types of superradiance, such as pure superradiance, triggered
superradiance, and pulsing superradiance.

The aim of the paper is to consider the realistic situation
corresponding to quantum dots, but not an oversimplified model
consideration. Therefore, in order to understand what
approximations are admissible for characterizing the process,
in the next section, an analysis of the parameters, typical
of semiconductors with quantum dots, will be given. Such an
analysis is necessary before plunging into mathematical
formulas in the following sections. The parameters are taken
from the literature cited above.

\section{Typical Material Parameters}

For realizing quantum-dot superradiance, it is reasonable to
take those materials that are used for quantum-dot lasers.
For the latter, one usually employs the self-assembled
heterostructures, such as InAs/GaAs,  InGaAs/GaAs,
InGaAs/AlGaAs, GaInAsP/InP, InAs/InP, InAs/GaInAs,
AlInAs/AlGaAs, InP/GaInP, AlGaAs/GaAs, and CdSe/ZnSe. There exist
different kinds of quantum dots, having different shapes and sizes. 
Thus, the lateral size of a typical self-assembled quantum dot is 
much larger than its vertical extent. Typical dot sizes are of the 
order $r_{dot}\sim 10^{-7} cm - 10^{-6} cm$. In each dot, there can
be between just a few to $10^5$ electrons. The dot density
in an epitaxy layer is of order $10^8 cm^{-2} - 10^{11} cm^{-2}$.
With the width of the layer $h \sim 10^{-6} cm - 10^{-5} cm$,
this makes the spatial density
$\rho \sim 10^{13} cm^{-3} - 10^{17} cm^{-3}$. The interdot distance
is $a \sim 10^{-5} cm - 10^{-4} cm$. The lasing operation is
realized at the wavelength $\lambda \sim 10^{-4} cm$, which
translates into the frequency $\omega_0 \sim 10^{15} Hz$. The
natural width $\gamma_0 = 2|d|^2 k_0^3/3$, where
$k_0 = 2\pi/\lambda$, depends on the transition dipole $d$.
The typical value for the latter is $d \sim 100 D$. Taking
into account that $1D = 10^{-17} \sqrt{erg\cdot cm^3}$ gives
$\gamma_0 \sim 10^{10} Hz$. The actual homogeneous broadening is
usually larger than $\gamma_0$, being
$\gamma_2 \sim 10^{12} Hz - 10^{13} Hz$. For high-quality
self-assembled dot materials, the inhomogeneous broadening can be
made relatively small, of the same order as $\gamma_2$, that is,
of order $\gamma_2^* \sim 10^{12} Hz - 10^{13} Hz$. The
longitudinal relaxation time $T_1$ is mainly due to
electron-phonon coupling, which can be strongly suppressed by
low temperatures. Thus, at helium temperatures, the longitudinal
dephasing time becomes limited only by the lifetime of
inversion for a single quantum dot in free space,
$T_1 \sim 10^{-9} s$, which gives
$\gamma_1 \equiv 1/T_1 \sim 10^9 Hz$.
To enhance a chosen mode, one places the sample into a resonator
cavity with a large quality factor reaching $10^4$. The sizes of
the sample can be different. Edge-emitting lasers do not have a 
circular cross-section. The typical sizes of quantum dot lasers
can be $R \sim 10^{-3} cm - 10^{-2} cm$.

Since the radiation wavelength is much larger than the dot sizes,
$\lambda \gg r_{dot}$, the dipole approximation is appropriate.
For sufficiently dense dot materials, the interdot distance can
be made much smaller than the wavelength, $a \ll \lambda$. Hence,
there can arise sufficiently strong dot interaction through the
common radiation field. But the wavelength is much smaller than
the sample linear sizes, $\lambda \ll L$. Therefore the
Dicke-type [36] approximation of a concentrated, point-like,
sample, cannot be used. In strongly nonuniform materials, with
a very large inhomogeneous broadening $\gamma_2^*$, such that
$T_2^* \equiv 1/\gamma_2^*$ is comparable with the time of the
radiation pulse $\tau_p$, superradiance is suppressed [34]. It
is therefore necessary to prepare the samples with a narrow
distribution of dot sizes, resulting in not too wide
inhomogeneous broadening. Fortunately, the fabrication of such
samples is nowadays technologically possible. As is seen from
the above values for typical parameters, the inhomogeneous
broadening can be made of the same order as the homogeneous one.
When $\gamma_2^* \sim \gamma_2$, then the consideration can be
simplified by combining $\gamma_2^*$ and $\gamma_2$ in one
effective parameter [37]. Thus, the typical dephasing time is
rather short, $T_2 \equiv 1/\gamma_2 \sim 10^{-13} s - 10^{-12} s$.
Superradiance is possible only if the time of radiation pulse
$\tau_p$ is shorter than the dephasing time $T_2$. In the
following sections, the theoretical description is developed,
which takes into account the typical characteristics of the
quantum dot assemblies. Throughout the paper, the system of units
is used, where the Planck constant $\hbar \equiv 1$.

\section{Basic Operator Equations}

Aiming at developing a realistic description of the system,
let us start with the microscopic Hamiltonian
\be
\label{1}
\hat H = \hat H_d + \hat H_f + \hat H_{df} +
\hat H_{mf} \; ,
\ee
characterizing an ensemble of radiating quantum dots in a
semiconductor matrix inside a resonator cavity. The Hamiltonian
\be
\label{2}
\hat H_d = \sum_{i=1}^N \om_0 \left ( \frac{1}{2} +
S_i^z \right ) \; ,
\ee
where $S_i^z$ is a pseudospin operator of an $i$-th dot,
represents $N$ two-level quantum dots, with the carrying
transition frequency $\omega_0$. Mathematically, the operator
$S_i^z$ is just a spin operator. It is called the pseudospin
operator, since it corresponds not to an actual spin but to
the population difference. The radiation-field Hamiltonian
\be
\label{3}
\hat H_f = \frac{1}{8\pi} \int \left (\bE^2 + \bH^2 \right ) \;
d\br \; ,
\ee
contains electric field $\bf E$ and magnetic field $\bf H$.
The vector potential $\bA$, introduced by the standard relation
${\bf H} = \nabla \times {\bf A}$, is assumed to satisfy the
Coulomb calibration $\nabla \cdot {\bf A} = 0$. The dot-field
interaction is given by the Hamiltonian
\be
\label{4}
\hat H_{df} = -\sum_{i=1}^N \left ( \frac{1}{c} \;
\bJ_i \cdot \bA_i + \bP_i \cdot \bE_{0i} \right ) \;  ,
\ee
in which ${\bf A}_i \equiv {\bf A}({\bf r}_i,t)$ and
${\bf E}_{0i} \equiv {\bf E}_0({\bf r}_i,t)$ is a seed
electric field of the resonator cavity. The transition
current
\be
\label{5}
\bJ_i = i\om_0 \left ( \bd S_i^+ - \bd^* S_i^- \right ) \; ,
\ee
and transition polarization
\be
\label{6}
\bP_i = \bd S_i^+ + \bd^* S_i^- \; ,
\ee
are expressed through the transition dipole $\bf d$ and the
ladder operators $S_i^\pm \equiv S_i^x \pm i S_i^y$, where
$S_i^\al \equiv S^\al({\bf r}_i,t)$. Since the cavity is filled 
by a semiconducting material, the Hamiltonian
\be
\label{7}
\hat H_{mf} = -\; \frac{1}{c} \int \bj_{mat} (\br,t) \cdot
\bA(\br,t) \; d\br  \; ,
\ee
describes the interaction of the local density current
$j_{mat}({\bf r},t)$ in the filling matter with the radiated
electromagnetic field.

The field operators satisfy the equal-time commutation
relations
$$
\left [ E^\al(\br,t), \; A^\bt(\br',t) \right ] =
4\pi i c \dlt_{\al\bt}(\br-\br') \; ,
$$
\be
\label{8}
\left [ E^\al(\br,t), \; H^\bt(\br',t) \right ] =
- 4\pi i c \sum_\gm \ep_{\al\bt\gm} \;
\frac{\prt}{\prt r^\gm} \; \dlt(\br-\br') \; ,
\ee
where $\epsilon_{\alpha \beta \gamma}$ is the unitary
antisymmetric tensor [3] and the transverse delta-function is
\be
\label{9}
\dlt_{\al\bt}(\br) \equiv \int \left (
\dlt_{\al\bt} \; - \; \frac{k^\al k^\bt}{k^2} \right )
e^{i\bk\cdot\br} \; \frac{d\bk}{(2\pi)^3} =
\frac{2}{3}\; \dlt_{\al\bt} \dlt(\br) \; - \;
\frac{1}{4\pi} \; D_{\al\bt}(\br) \;  ,
\ee
with the dipolar tensor
$$
D_{\al\bt}(\br) \equiv
\frac{\dlt_{\al\bt} - 3 n^\al n^\bt}{r^3} \;  ,
$$
in which ${\bf n} \equiv {\bf r}/r = \{n^\alpha\}$ and
$r \equiv |\bf r|$. The pseudospin operators satisfy the spin
commutation relations
\be
\label{10}
\left [ S^+_i, \; S_j^- \right ] = 2 \dlt_{ij} S_i^z \; ,
\qquad \left [ S^z_i, \; S_j^\pm \right ] =
\pm \dlt_{ij} S_i^\pm  \; .
\ee

The Heisenberg equations of motion for the field operators
yield
\be
\label{11}
\frac{1}{c} \; \frac{\prt \bE}{\prt t} =
\nabla \times \bH \; - \; \frac{4\pi}{c}\; \bj  \; ,
\qquad \frac{1}{c} \; \frac{\prt\bA}{\prt t} = - \bE \; ,
\ee
with the density of current
\be
\label{12}
j^\al(\br,t) = \sum_\bt \left [
\sum_{i=1}^N \dlt_{\al\bt}(\br-\br_i) J_i^\bt(t) +
\int \dlt_{\al\bt}(\br-\br') j_{mat}^\bt (\br',t)\; d\br'
\right ]  \; .
\ee
>From these, using the Coulomb calibration, one gets the
equation for the vector potential
\be
\label{13}
\left ( \nabla^2 \; - \; \frac{1}{c^2} \;
\frac{\prt^2}{\prt t^2} \right ) \bA = - \;
\frac{4\pi}{c} \; \bj \;  .
\ee

The Heisenberg equations for the pseudospin operators result
in the equations
$$
\frac{dS_i^-}{dt} = - i\om_0 S_i^- + 2 S_i^z \left (
k_0 \bd \cdot \bA_i - i \bd \cdot \bE_{0i} \right ) \; ,
$$
\be
\label{14}
\frac{d S_i^z}{dt} = - S_i^+ \left ( k_0 \bd \cdot \bA_i -
i \bd \cdot \bE_{0i} \right )  - S_i^-
\left ( k_0 \bd^* \cdot \bA_i +  i \bd^* \cdot \bE_{0i}
\right )\; ,
\ee
where $k_0 = \omega_0/c$. These equations are to be
complimented by the retardation condition
\be
\label{15}
S_i^\al (t) = 0 \qquad (t < 0) \; .
\ee

Equations (11) to (15) are the basic operator equations
describing all radiation processes in the system of quantum
dots. It is worth emphasizing that these equations follow
from the first principles. Such a microscopic approach is
necessary for correctly treating the overall dynamics of
quantum dot radiation.

\section{Elimination of Field Variables}

The standard way of considering the radiation processes is
by averaging Eqs. (11) and (14) and passing to the
semiclassical approximation. This way, presupposing well
organized coherence, does not allow for the treatment of those
radiation stages, when coherence has not been developed.
Therefore, we employ here another, more accurate, approach
allowing for the treatment of all radiation stages.

The first step in the approach to be pursued is the elimination
of field variables [8]. This can be done by solving Eq. (13)
for the vector potential and substituting the found solution into
Eqs. (14) for the pseudospin operators. The known solution of
Eq. (13) is the sum
\be
\label{16}
\bA(\br,t) = \bA_{vac} + \frac{1}{c} \int \bj \left ( \br',
t - \; \frac{|\br-\br'|}{c} \right ) \; \frac{d\br'}{|\br-\br'|}
\ee
of the vacuum potential and the retarded potential, with the
density of current given by Eq. (12). This solution is to be
substituted into Eqs. (14).

The interaction between the radiation field and a radiating dot
is assumed to be small as compared to the transition frequency.
This is a necessary requirement for the existence of well defined
energy levels in a dot. In the other case, the transition
frequency would not be defined in principle. This implies that
the retardation in the time dependence of the current density
can be taken into account in the Born approximation
$$
S_j^- \left (t-\frac{r}{c}\right ) =
S_j^-(t)\Theta(ct-r) e^{ik_0r} \; , \qquad
S_j^z \left (t-\frac{r}{c}\right ) = S_j^z(t)\Theta(ct-r) \; ,
$$
where $\Theta(t)$ is the unit step function.

When substituting the vector potential (16) into Eqs. (14), one
meets the terms corresponding to the dot self-action. The 
contribution of these terms is characterized in the Appendix.
Finally, the vector potential (16) is represented as the sum
\be
\label{17}
\bA = \bA_{vac} + \bA_{self} + \bA_{rad} + \bA_{dip} +
\bA_{mat}\;  .
\ee
Here the first term is due to vacuum fluctuations. The self-action
potential is described in the Appendix. The radiation potential
\be
\label{18}
\bA_{rad}(\br,t) = \sum_j \; \frac{2}{3c|\br-\br_j|} \;
\bJ_j \left ( t - \; \frac{|\br-\br_j|}{c} \right ) \; ,
\ee
in which ${\bf r}\neq {\bf r}_j$, describes the spherical part 
of the vector potential, corresponding to the radiation field
produced by dots. In addition, the radiating dots create the
dipolar part of the vector potential
\be
\label{19}
A_{dip}^\al(\br,t) =  -\sum_j \sum_\bt \int
\frac{D_{\al\bt}(\br'-\br_j)}{4\pi c|\br-\br'|} \;
J_j^\bt \left ( t - \; \frac{|\br-\br'|}{c}
\right ) \; d\br' \;  .
\ee
The interaction of the semiconductor, filling the cavity, with the
radiation field produces the potential
\be
\label{20}
A_{mat}^\al(\br,t) = \sum_\bt \int
\frac{\dlt_{\al\bt}(\br'-\br'')}{c|\br-\br'|} \;
j_{mat}^\bt \left (\br'', t - \; \frac{|\br-\br'|}{c}
\right ) \; d\br' d\br''\;   .
\ee
Substituting the vector potential (17) into Eqs. (14), we obtain
the equations for the pseudospin operators, which do not contain
the field variables. Instead, there appear effective dot
interactions through the common radiation field.

\section{Stochastic Mean-Field Approximation}

The dynamics of the system can be characterized by the behavior
of the following functions. The {\it transition function}
\be
\label{21} 
u(\br_i,t) \equiv 2 \lgl S_i^-(t) \rgl_H
\ee
where the angle brackets imply quantum statistical averaging
associated with the system Hamiltonian $H$, describes
the local effective polarization corresponding to dipole
transitions. The {\it coherence intensity}
\be
\label{22}
w(\br_i,t) \equiv \frac{2}{N} \; \sum_{j(\neq i)}^N
\left [ \lgl S_i^+(t) S_j^-(t) \rgl_H +
\lgl S_j^+(t) S_i^-(t) \rgl_H \right ]
\ee
is the local characteristic of coherence. The intensity of
coherent radiation is proportional to this function. The
{\it population difference}
\be
\label{23}
s(\br_i,t) \equiv 2 \lgl S_i^z(t) \rgl_H
\ee
defines the local difference of populations for the energy
levels of an $i$-dot.

To obtain the evolution equations for functions (21), (22),
and (23), we have to average the equations resulting after
the substitution of the vector potential (17) into the
equations of motion (14). Such equations are not closed.
To make them closed, it is necessary to invoke some
decoupling for the operator correlation functions. If one
resorts to the standard mean-field decoupling, one comes to
the usual semiclassical approximation. As is well known, the
semiclassical approximation can be used only when the system
is in a coherent state or is almost coherent [2,3]. But
incoherent regimes cannot be described in this approximation.
One of the most interesting questions is how coherence
develops in an initially incoherent system. To be able to
describe such a regime, it is necessary to employ a more
accurate approximation. For this purpose, we shall use the
{\it stochastic mean-field approximation} employed earlier
for describing the dynamics of spin assemblies [5-11] and
Bose systems in random fields [38-40].

Let us combine the vacuum, dipole, and matter vector potentials
into the sum
\be
\label{24}
\xi(\br,t) \equiv 2k_0\bd \cdot \left ( \bA_{vac} +
\bA_{dip} + \bA_{mat} \right ) \;  .
\ee
The potentials, entering this sum, create local fluctuations of
electromagnetic field. Being averaged over space, quantity (24)
is practically zero. Therefore Eq. (24), characterizing the
strength of local field fluctuations, can be treated as a
local random variable. Contrary to this random variable, the
radiation potential (18) induces long-range effective
interactions between the radiating dots. The long-range
radiation potential (18), with current (5), is expressed through
the pseudospin operators ${\bf S}_j$. Hence the effective
interactions, arising between dots, are mathematically similar
to long-range spin interactions.

There is a direct similarity between the effective dipole
interactions caused by the dipole vector potential (19) and the
dipole [5-11] and hyperfine interactions [9-11,41,42] in spin
systems, these interactions being treated as stochastic
fluctuations playing destructive role by dephasing collective
motion. While the effective pseudospin interactions, induced by
the radiation potential (18) in quantum dots, are equivalent to
the effective interactions produced by the resonator feedback
field in spin systems [9-11]. The latter interactions are
responsible for the appearance of collective effects in magnetic
and ferroelectric samples, including the arising coherence [43].
In spin systems without a resonator feedback field, coherence
and, hence, superradiance, cannot develop, being destroyed by
the direct dipole interactions [5-11,44,45].

In this way, it is possible to distinguish two types of variables
in the system, the random variable $\xi$ describing local field
fluctuations and the pseudospin variables ${\bf S}_j$
characterizing effective long-range dot interactions. For brevity,
we may denote the collection of pseudospins
$\{ \bS_j: j = 1,2,...,N\}$ by $\bf S$. Then, the operators of
observable quantities $\hat{\cal O}$ are, generally, functions of
these variables, $\hat{\cal O} = \hat{\cal O}(\bS,\xi)$.

Having two kinds of variables, it is possible to define two types
of averaging. One type corresponds to the {\it quantum statistical}
average
\be
\label{25}
\lgl \hat{\cal O}\rgl_H \equiv
{\rm Tr} \;\hat\rho \; \hat{\cal O}(\bS,\xi)
\ee
involving the pseudospin operators $\bf S$, with $\hat{\rho}$
being the system statistical operator, and where the random
variable $\xi$ is kept fixed. Another type is the {\it stochastic}
averaging over the random fluctuations, denoted as
\be
\label{26}
\lgl \lgl \hat{\cal O} \rgl \rgl \equiv
\int \hat\rho\; \hat{\cal O}(\bS,\xi) \; {\cal D}\xi \;  ,
\ee
which is defined through a functional integral over the random
variable $\xi$, with the prescribed differential measure
${\cal D}\xi$.  Respectively, the total averaging
\be
\label{27}
\lgl \hat{\cal O} \rgl \equiv \lgl \lgl \left (
\lgl \hat{\cal O} \rgl_H \right ) \rgl\rgl
\ee
includes both, the quantum and stochastic, averages.

Keeping in mind the long range of the effective pseudospin
interactions, we can use the {\it stochastic mean-field
decoupling}
\be
\label{28}
\lgl S_i^\al S_j^\bt \rgl_H = \lgl S_i^\al \rgl_H
\lgl S_j^\bt \rgl_H \qquad (i\neq j) \;  ,
\ee
where only the quantum averaging is involved. This decoupling
looks like a mean-field approximation. However, it has a very
important principal difference form the latter, involving only
the quantum averaging (25), but not touching the stochastic
averaging (26). Since no approximation is done here with respect
to the stochastic variables, decoupling (28) preserves stochastic
properties of the system. This is why it is called the
stochastic mean-field approximation [5-11].

When the radiation wavelength $\lambda$ is much larger than the
interdot distance, the geometrical location of dots in space is
of no importance. Then it is convenient to pass to the continuous
spatial representation, replacing the sums by the integrals
according to the rule
\be
\label{29}
\sum_{j=1}^N \; \Longrightarrow \; \rho \int d\br
\qquad \left ( \rho \equiv \frac{N}{V} \right ) \;  ,
\ee
with the integration over the whole system and $\rho$ being the
dot density.

In order to represent the evolution equations in a compact form,
let us introduce the effective field, or effective force, acting
on dots
\be
\label{30}
f(\br,t) =  f_0(\br,t) + f_{rad}(\br,t) +
\xi(\br,t) \;  .
\ee
Here, the first term
\be
\label{31}
f_0(\br,t) \equiv -2 i \bd \cdot \bE_0(\br,t)
\ee
is due to the external field ${\bf E}_0$. The second term
\be
\label{32}
f_{rad}(\br,t) \equiv 2 k_0 \lgl \bd \cdot \bA_{rad}(\br,t)
\rgl_H
\ee
is caused by the radiating dots. And the last term in Eq. (30)
is the fluctuating random field (24). The radiation term (32),
with the vector potential (18), acquires the form
\be
\label{33}
f_{rad}(\br,t) = - i\gm_0 \rho \int \left [
G(\br-\br',t) u(\br',t) - \; \frac{\bd^2}{|\bd|^2} \;
G^*(\br-\br',t) u^*(\br',t) \right ] \; d\br' \; ,
\ee
in which the transfer kernel is
\be
\label{34}
G(\br,t) \equiv \frac{\exp(ik_0 r)}{k_0r} \;
\Theta(ct-r) \;  .
\ee

Finally, we average Eqs. (14) according to the quantum
averaging (25), employ the stochastic mean-field decoupling
(28), and use notation (30). Then for the variables (21),
(22), and (23), we obtain the evolution equations
$$
\frac{\prt u}{\prt t} = - (i\om_0 + \gm_2) u + f s \; ,
\qquad
\frac{\prt w}{\prt t} = - 2\gm_2 w +
\left ( u^* f + f^* u \right ) s \; ,
$$
\be
\label{35}
\frac{\prt s}{\prt t} = -\; \frac{1}{2}
\left ( u^* f + f^* u \right ) - \gm_1 ( s -\zeta )\; ,
\ee
in which $u = u({\bf r},t), w = w({\bf r},t)$, and
$s = s({\bf r},t)$. The longitudinal, $\gamma_1$, and
transverse, $\gamma_2$, attenuation rates are treated as the
system parameters, whose typical values are discussed in
Sec. II. The parameter $\zeta$ characterizes the level of
stationary nonresonant pumping. The evolution equation for $w$
follows from definition (22), with the use of the Heisenberg 
equations of motion for the pseudospin operators and the 
stochastic mean-field decoupling (28). Since in definition (22),
the summation is over $j \neq i$, it is also possible, first,
to decouple the products of the pseudospin operators, according
to Eq. (28), and then invoke the Heisenberg equations. In any 
case, the result is the same equation for $w$.   

Equations (35) are stochastic integro-differential equations. 
These are the basic evolution equations describing the dynamics 
of radiation in a system of quantum dots.

\section{Triggering Dipolar Waves}

As is mentioned above, the dipolar vector potential (19)
induces local fluctuations dephasing the radiation of the
dot assembly. However, these fluctuations play not only
destructive role. They can be useful at the initial stage of
the radiation process, when the latter is not triggered by
an external field. In such a case, the transition dipole
fluctuations, related to spontaneous emission, can trigger
the process of collective radiation.

In order to illustrate the appearance and the nature of the
dipolar fluctuations, let us consider the pseudospin
equations (14), leaving there only the terms related to the
dipole interactions. Let us introduce the interaction
coefficients
\be
\label{36}
b_{ij} \equiv \frac{k_0^2}{2\pi} \;
\sum_{\al\bt} d^\al D_{ij}^{\al\bt} d^\bt \; , \qquad
c_{ij} \equiv \frac{k_0^2}{2\pi} \;  \sum_{\al\bt}
d^\al \left (D_{ij}^{\al\bt} d^\bt \right )^* \;  ,
\ee
where
$$
D_{ij}^{\al\bt} \equiv \int \Theta(ct - |\br_i -\br'| )\;
\frac{D_{\al\bt}(\br'-\br_j)}{|\br_i-\br'|}
\exp(-ik_0|\br_i-\br'|) \; d\br' \; .
$$
The latter quantity, because of the unit-step function
$\Theta(ct-|\br_i-\br'|)$ in the integral, varies in time
only at the very beginning of the process, when $t\ll a/c$, 
after which it becomes practically constant. Therefore, for 
all times, except $t\ll a/c$, the interaction coefficients
(36) can be treated as constant parameters.

For the dipolar vector potential (19), we have
$$
k_0 \bd \cdot\bA_{dip}(\br_i,t) = - \;
\frac{i}{2} \sum_{j(\neq i)} \left [ b_{ij} S_j^+(t) -
c_{ij} S_i^-(t) \right ] \; .
$$

Then the equations of motion reduce to
$$
\frac{dS_i^-}{dt} = - i\om_0 S_i^- - i S_i^z
\sum_{j(\neq i)} \left ( b_{ij} S_j^+ - c_{ij} S_j^-
\right ) \; ,
$$
\be
\label{37}
 \frac{dS_i^z}{dt} = \frac{i}{2} \; \sum_{j(\neq i)}
\left [ S_i^+ \left ( b_{ij} S_j^+ - c_{ij} S_j^-\right ) -
S_i^- \left ( b_{ij}^* S_j^- - c_{ij}^* S_j^+ \right )
\right ] \;  .
\ee
The dipolar pseudospin fluctuations are described by the small
deviations
\be
\label{38}
\dlt S_i^\pm \equiv S_i^\pm - \lgl S_i^\pm \rgl \; , \qquad
\dlt S_i^z \equiv S_i^z - \lgl S_i^z \rgl
\ee
from the related average values $<S_i^{\alpha}>$. The latter
averages correspond to equilibrium or quasiequilibrium, when
they either do not depend on time or are slow functions of time,
as compared to the fastly fluctuating deviations (38).
Linearizing Eqs. (37) with respect to small deviations (38),
under the condition $<S_i^-> = 0$, and taking into account that,
because of the properties of dipole interactions,
\be
\label{39}
\sum_{j(\neq i)} b_{ij} = \sum_{j(\neq i)} c_{ij} = 0 \; ,
\ee
we obtain the equations 
$$
\frac{d}{dt} \; \dlt S_i^- = - i\om_0 \dlt S_i^- -
i\lgl S_i^z \rgl \sum_{j(\neq i)} \left ( b_{ij} \dlt S_j^+ -
c_{ij} \dlt S_j^- \right ) \; ,
$$
\be
\label{40}
\frac{d}{dt} \; \dlt S_i^z = 0\; .
\ee

Then we employ the Fourier transforms for the pseudospin
operators
\be
\label{41}
S_j^\pm = \sum_k S_k^\pm \exp ( \mp i \bk \cdot \br_j) \;  ,
\qquad S_k^\pm = \frac{1}{N} \sum_j S_j^\pm \exp(\pm i\bk \cdot
\br_j) \; ,
\ee
for the interaction coefficients
\be
\label{42}
b_{ij} = \frac{1}{N} \sum_k b_k
\exp (i\bk \cdot \br_{ij} ) \; ,\qquad
b_k = \sum_{j(\neq i)} b_{ij}
\exp(-i\bk \cdot \br_{ij}) \; ,
\ee
and, similarly, for the interaction coefficients $c_{ij}$,
where ${\bf r}_{ij} \equiv {\bf r}_i - {\bf r}_j$. Taking
into account that
\be
\label{43}
\dlt S_j^\pm = S_j^\pm \qquad ( \lgl S_j^\pm \rgl = 0 )
\ee
and introducing the notation
\be
\label{44}
\mu_k \equiv \om_0 - c_k \lgl S_i^z \rgl \; , \qquad
\lbd_k \equiv b_{-k} \lgl S_i^z \rgl \; ,
\ee
we come to the equations
\be
\label{45}
\frac{d S_k^-}{dt} = -i\mu_k S_k^- - i\lbd_k S_k^+ \; ,
\qquad \frac{d S_k^+}{dt} = i\mu_k^* S_k^+ + i\lbd_k^* S_k^- \;  .
\ee

Looking for the solutions to Eqs. (45) in the form
\be
\label{46}
S_k^- = u_k e^{-i\om_k t} + v_k^* e^{i\om_k t}  \; ,
\qquad S_k^+ = u_k^* e^{i\om_k t} + v_k e^{-i\om_kt} \; ,
\ee
we find the dipolar-wave spectrum
\be
\label{47}
\om_k = \sqrt{|\mu_k|^2 - |\lbd_k|^2} \; .
\ee
This means that the dipolar part (19) of the vector potential
generates local field fluctuations, having the meaning of the
transition dipolar waves with the spectrum (47). The
interaction coefficients (36) are smaller than the transition
frequency, so that
\be
\label{48}
\frac{|b_k|}{\om_0} \ll 1 \; , \qquad
\frac{|c_k|}{\om_0} \ll 1  \;   .
\ee
Hence,
\be
\label{49}
\frac{|\lbd_k|}{\om_0} \ll 1 \; , \qquad
\frac{|\lbd_k|}{|\mu_k|} \ll 1  \;   .
\ee
Therefore the spectrum (47) is always positive, implying that
the dipolar waves are dynamically stable [46]. In the long-wave
limit, when $k \ra 0$ and
$$
b_k \simeq -\; \frac{1}{2} \sum_{j(\neq i)} b_{ij}
(\bk \cdot \br_{ij})^2 \; , \qquad
c_k \simeq -\; \frac{1}{2} \sum_{j(\neq i)} c_{ij}
(\bk \cdot \br_{ij})^2 \; ,
$$
the spectrum becomes quadratic, which follows from the
expression
\be
\label{50}
\om_k^2 \simeq \om_0^2 + \om_0 \lgl S_i^z \rgl\; {\rm Re}
\sum_{j(\neq i)} c_{ij} (\bk\cdot \br_{ij})^2 \; .
\ee
The existence of these dipolar waves triggers the process of
dot radiation, even if no external field is imposed at the
initial time.

\section{Transverse Mode Expansion}

After the radiation process is triggered by the dipolar waves,
the overall radiation dynamics is described by Eqs. (35) that
are stochastic integro-differential equations in partial
derivatives. If the sizes of the whole sample would be much
smaller than the radiation wavelength, we could essentially
simplify the problem by resorting to the concentrated-sample
approximation [1-3,36], when just one mode fills the cavity.
But in realistic situation, vice versa, the wavelength is
usually much smaller than the sample sizes, so that the sample
can house several modes. Then, to simplify the equations, it is
necessary to specify the shape of the resonator cavity.

For concreteness, let us assume that the cavity is cylindrical,
with radius $R$ and length $L$, such that the wavelength is much
smaller than these sizes:
\be
\label{51}
\frac{\lbd}{R} \ll 1 \; , \qquad
\frac{\lbd}{L} \ll 1 \; .
\ee
Directing the cylinder axis along the axis $z$, we can treat
the resonator seed field as propagating along this axis, that is,
having the form
\be
\label{52}
E_0(\br,t) = \frac{1}{2}\; E_1 e^{i(kz-\om t)} +
\frac{1}{2} \; E_1^* e^{-i(kz-\om t)} \; .
\ee
The cavity is resonant in the sense of the small detuning of
the resonator natural frequency from the dot transition
frequency,
\be
\label{53}
\frac{|\Dlt|}{\om_0} \ll 1 \qquad (\Dlt \equiv
\om - \om_0) \;  .
\ee
In the same way as it is for all dot-field interactions, for
the seed field, we have
\be
\label{54}
\frac{|\nu_1|}{\om_0} \ll 1 \qquad (\nu_1 \equiv |\bd \cdot
\bE_1| ) \;  .
\ee

In the cylindrical geometry, the radiation modes acquire the
shape of filaments extended along the axis $z$. Then, it is
possible to represent the solutions to Eqs. (35) as expansions
over the transverse modes:
$$
u(\br,t) = \sum_{n=1}^{N_f} u_n(r_\perp,t) e^{ikz}\;  , \qquad
w(\br,t) = \sum_{n=1}^{N_f} w_n(r_\perp,t) \;  ,
$$
\be
\label{55}
s(\br,t) = \sum_{n=1}^{N_f} s_n(r_\perp,t) \;  ,
\ee
where $N_f$ is the number of the filamentary modes and
$r_{\perp} = \sqrt{x^2 + y^2}$ is the transverse radial
variable. Ascribing to a mode an effective enveloping
radius $R_f$, for the effective enveloping volume of a
filamentary mode, we have $V_f = \pi R_f^2 L$. Representation
(55) is rather general, including the case of just a single
mode, when $N_f = 1$ and $V_f = V$, with $V$ being the sample
volume. We keep in mind that the sample and cavity volumes
coincide. This assumption does not reduce the generality of
consideration, since when these volumes are different it is
sufficient to take into account the existence of a filling
factor that is not equal to one.

In what follows, we assume that the distance $d_f$ between 
the axes of any two nearest-neighbor filaments is larger than 
twice the filament enveloping radius. The physical condition, 
corresponding to this assumption can be understood by taking 
into account that the filament enveloping radius is of the order 
of the diffraction radius, i.e., $R_f \sim \sqrt{\lambda L}$.
Then, from the inequalities $R_f \ll d_f < R$, it follows that 
the Fresnel number $F = R^2/ \lambda L$ is to be large, $F \gg 1$.
Therefore, in the case of large Fresnel numbers, different 
filamentary modes can be treated as uncorrelated and considered 
separately from each other. In such a case, we can reduce the 
consideration to studying the behavior of each mode on average 
by defining the averages over the enveloping volume of a mode as
$$
u(t) \equiv \frac{1}{V_f} \int_{V_f} u_n(r_\perp,t) \; d\br =
\frac{2}{R_f^2} \int_0^{R_f} u_n(r,t) r dr \; ,
$$
$$
w(t) \equiv \frac{1}{V_f} \int_{V_f} w_n(r_\perp,t) \; d\br =
\frac{2}{R_f^2} \int_0^{R_f} w_n(r,t) r dr \; ,
$$
\be
\label{56}
s(t) \equiv \frac{1}{V_f} \int_{V_f} s_n(r_\perp,t) \; d\br =
\frac{2}{R_f^2} \int_0^{R_f} s_n(r,t) r dr \; ,
\ee
where, for the simplicity of the following notation, the index
$n$ enumerating modes is omitted in the left-hand side of these
equations.

We also need to introduce the coupling functions
$$
\al(t) \equiv \gm_0 \rho \int_{V_f} \Theta(ct-r) \;
\frac{\sin(k_0r - kz)}{k_0 r} \; d\br  \; ,
$$
\be
\label{57}
\bt(t) \equiv \gm_0 \rho \int_{V_f} \Theta(ct-r) \;
\frac{\cos(k_0r - kz)}{k_0 r} \; d\br  \; ,
\ee
the average stochastic field
\be
\label{58}
\xi(t) \equiv \frac{1}{V_f} \int \xi(\br,t) e^{-ikz}\; d\br \; ,
\ee
and the effective force
\be
\label{59}
f_1(t) \equiv - i \bd \cdot \bE_1 e^{-i\om t} + \xi(t) \; .
\ee
Then, we substitute the mode expansions (55) into Eqs. (35), 
averaging the mode functions according to Eqs. (56). We use the
condition that different spatial modes are not correlated with 
each other, so that
$$
\sum_{mn} u_m(r_\perp,t) e^{ikz} s_n(r_\perp,t) =
\sum_n u_n(r_\perp,t) e^{ikz} s_n(r_\perp,t) \;   .
$$
And we employ the theorem of average in the integral
$$
\int_{V_f} G(\br' -\br,t) u_n(r_\perp,t) e^{ikz} 
s_n (r_\perp',t) \; d\br d\br' = u(t) s(t) \int_{V_f}
G(\br'-\br,t) e^{ikz}\; d\br d\br' \;   .
$$ 
Thus, using the above notations (57), (58), and (59), we obtain 
the equations
$$
\frac{du}{dt} = - i(\om_0 +\bt s) u -
(\gm_2 - \al s) u + f_1 s \; ,
$$
$$
\frac{dw}{dt} = - 2(\gm_2 - \al s) w +
\left ( u^* f_1 + f_1^* u \right ) s \; ,
$$
\be
\label{60}
 \frac{ds}{dt} = -\al w  - \; \frac{1}{2}
\left ( u^* f_1 + f_1^* u \right ) s - \gm_1 (s -\zeta)\; ,
\ee
describing the evolution of an averaged mode
characterized by functions (56).

In this way, the mode expansions (55) allow us to
transform Eqs. (35) in partial derivatives to Eq. (60)
in ordinary derivatives. The used mode-expansion method
is based on the idea of the eikonal approximation [47,48].

Formally, the used mode-expansion method reduces the general 
problem to a collection of effective single-mode radiation 
problems, each decoupled from the others. Such a reduction is, 
actually, the main idea of the eikonal approximation. Greatly 
simplifying the consideration, this reduction leaves aside the 
question of what would be the distribution of filament sizes.
The latter is a separate problem, depending on the sample shape.

\section{Scale Separation Approach}

Equations (60) are stochastic differential equations that
are not easy to solve. Fortunately, they can be further
simplified using the scale separation approach [5-8,43]
that is a variant of the averaging technique [49]. In the
present section, we employ this approach [5-8,43].

It is possible to notice that there are different time
scales in this system of equations. The attenuation rates,
as discussed in Sec. II, are small as compared to the
transition frequency, thus, defining the small parameters
\be
\label{61}
\frac{\gm_0}{\om_0} \ll 1 \; , \qquad
\frac{\gm_1}{\om_0} \ll 1 \; , \qquad
\frac{\gm_2}{\om_0} \ll 1 \; .
\ee
It, therefore, follows from Eqs. (60) that the function
$u(t)$ is fast in time, as compared to the slow functions
in time $w(t)$ and $s(t)$. It is convenient to introduce
other slow functions having the meaning of the
{\it collective width}
\be
\label{62}
\Gm \equiv \gm_2 - \al s \; ,
\ee
{\it collective frequency}
\be
\label{63}
\Om \equiv \om_0 + \bt s \; ,
\ee
and {\it effective detuning}
\be
\label{64}
\dlt \equiv \om - \Om = \Dlt - \bt s \; .
\ee
These slow functions play the role of quasi-integrals of
motion, or quasi-invariants, for the fast function $u(t)$.
The first of Eqs. (60) can be solved by keeping fixed the
quasi-integrals of motion, which gives
\be
\label{65}
u = \left ( u_0 - \; \frac{\nu_1 s}{\dlt + i\Gm}
\right ) e^{-(i\Om+\Gm)t} + \frac{\nu_1 s}{\dlt +
i\Gm} e^{-i\om t} + s \int_0^t \xi(t') e^{-(i\Om+\Gm)(t-t')}
\; dt' \; .
\ee
The seed field (52) is defined up to a phase factor. Then,
without the loss of generality, the global phase of the
seed field can be chosen such that the value
$u_0 {\bf d} \cdot {\bf E}_1$ be real, where $u_0 \equiv u(0)$.

The found solution (65) has to be substituted into the
second and third of Eqs. (60) for the slow functions $w(t)$
and $s(t)$. The right-hand sides of the latter equations are
to be averaged over the explicitly entering time and over
the stochastic variable according to the rule
\be
\label{66}
\lim_{\tau\ra\infty} \; \frac{1}{\tau}
\int_0^\tau \lgl \lgl \ldots \rgl \rgl dt \; ,
\ee
with keeping fixed the quasi-invariants. The stochastic
variable $\xi$, describing local field fluctuations, by its
definition, is zero-centered, such that
\be
\label{67}
\lgl \lgl \xi(t) \rgl \rgl = 0 \; .
\ee
And the correlation function $\ll \xi^*(t) \xi(t') \gg$
defines the {\it dynamic attenuation rate}
\be
\label{68}
\gm_3 \equiv {\rm Re} \lim_{\tau\ra \infty} \;
\frac{1}{\tau} \int_0^\tau dt \;
\int_0^t \lgl\lgl \xi^*(t) \xi(t')\rgl\rgl
e^{-(i\Om+\Gm)(t-t')} \; dt' \; ,
\ee
caused by these random fluctuations. The dynamic attenuation 
rate (68) is essentially defined by the currents in the 
semiconductor sample. These currents are usually much stronger
than those fluctuating in free space. This fact makes the 
principal difference between the considered case of quantum dots 
in semiconductor and atoms in free space. For the latter, the 
attenuation rate (68) is usually much smaller than the transverse
relaxation rate $\gamma_2$, while for semiconductors, on the 
contrary, $\gamma_3 \geq \gamma_2$. 

Let us introduce the {\it effective attenuation rate}
\be
\label{69}
\Gm_3 \equiv \gm_3 + \frac{|\nu_1|^2\Gm}{\dlt^2+\Gm^2}
\left ( 1 - e^{-\Gm t} \right ) \; ,
\ee
where $|\delta| < |\Gamma|$. Following the described
averaging procedure, we obtain the equations for the
{\it guiding centers}
\be
\label{70}
\frac{dw}{dt} = - 2 (\gm_2 -\al s) w + 2 \Gm_3 s^2 \; ,
\qquad
\frac{ds}{dt} = -\al w - \Gm_3 s - \gm_1 ( s -\zeta) \; .
\ee
Thus, the fast variables have been averaged out, while
the derived Eqs. (70) characterize the evolution of the
slow variables.

\section{Dynamics of Dot Radiation}

The temporal evolution of dot radiation through transverse 
modes is described by Eqs. (70). The solutions to these equations
essentially depend on the behavior of the coupling functions (57), 
whose values vary with time. It is possible to distinguish several 
qualitatively different stages of evolution.

\vskip 5mm
{\bf A. Fluctuation Stage}

\vskip 2mm

At the initial time $t = 0$, the coupling functions (57) are
zero. They remain small during the time interval
\be
\label{71}
0 < t < t_{int} \; ,
\ee
before the {\it interaction time} $t_{int}=a/c$, when the 
dots have not yet been correlated by means of the photon 
exchange. The interaction time, for the interdot distance
$a \sim 10^{-5} cm - 10^{-4} cm$ is 
$t_{int} \sim 10^{-15} s - 10^{-14} s$. In the time interval
(71), the radiation process starts, being triggered by dipolar
waves considered in Sec. VI. These waves correspond to the
random local field fluctuations. As is seen from the above
estimates, the fluctuation stage is rather short. During this
stage, the functions $w(t)$ and $s(t)$ do not essentially
change, so that $w(t_{int}) \approx w(0)$ and
$s(t_{int}) \approx s(0)$.

\vskip 5mm
{\bf B. Quantum Stage}

\vskip 2mm

The quantum stage comes after the time $t_{int}$, when the dot
interactions through photon exchange come into play, but dots
are not yet sufficiently correlated in order that coherence
would develop. At this incoherent stage, dots radiate
independently. The stage lasts till the {\it coherence time}
$t_{coh}$ that is necessary for developing coherence. So, the
temporal interval related to the quantum stage is
\be
\label{72}
t_{int} < t < t_{coh} \; .
\ee
During the interaction time $t_{int}$, the coupling functions
(57) quickly grow, reaching, after $t_{int}$, their maximal
values:
$$
\al(t) \ra g \gm_2 \; , \qquad
\bt(t) \ra \tilde g \gm_2 \qquad (t > t_{int}) \; .
$$
Here we have introduced the dimensionless {\it coupling
parameters}
\be
\label{73}
g \equiv \rho \; \frac{\gm_0}{\gm_2} \; \int_{V_f}
\frac{\sin(k_0r-kz)}{k_0r} \; d\br
\ee
and, respectively,
\be
\label{74}
\tilde g \equiv \rho \; \frac{\gm_0}{\gm_2} \; \int_{V_f}
\frac{\cos(k_0r-kz)}{k_0r} \; d\br \; .
\ee

At this stage, the collective width (62) becomes
\be
\label{75}
\Gm = \gm_2 ( 1 - g s )
\ee
and the collective frequency (63) is
\be
\label{76}
\Om = \om_0 + \tilde g \gm_2 s  \; .
\ee
The role of the resonator seed field (52) is to select
the resonant frequency, but its amplitude is small, such
that $|\nu_1| \ll \gamma_2$. Therefore the effective
attenuation (69) simplifies to $\Gamma_3 \simeq \gamma_3$.

Using the above expressions and considering the case when
at the initial time no coherence is imposed by external
fields, so that $w_0 \equiv w(0) = 0$, from Eqs. (70) we
have the equations
\be
\label{77}
\frac{dw}{dt} = 2\gm_3 s^2 \; , \qquad
\frac{ds}{dt} = - (\gm_1 + \gm_3) s + \gm_1 \zeta\;  .
\ee
The second of these equations yields the population
difference
\be
\label{78}
s = \left ( s_0 - \; \frac{\gm_1\zeta}{\gm_1 + \gm_3} 
\right ) \; \exp \{ - (\gm_1 + \gm_3)t \}\; + \;
\frac{\gm_1\zeta}{\gm_1+\gm_3} \; .
\ee
At short times, when $(\gamma_1 + \gamma_3)t \ll 1$, the
population difference is
\be
\label{79}
s \simeq s_0 - \left ( s_0 - 
\frac{\gm_1\zeta}{\gm_1 + \gm_3} \right ) (\gm_1 + \gm_3)t \; + \;
\frac{1}{2} \left ( s_0 - \frac{\gm_1\zeta}{\gm_1 + \gm_3} 
\right ) (\gm_1 + \gm_3)^2t^2 \; .
\ee
Then the coherence intensity behaves as
$$
w \simeq 2\gm_3 s_0^2 t - 2\gm_3 s_0 ({\gm_1 + \gm_3} )
\left ( s_0 - \frac{\gm_1\zeta}{\gm_1 + \gm_3} 
\right ) t^2 +
$$ 
\be
\label{80}
+ \frac{2}{3}\; \gm_3 ({\gm_1 + \gm_3} )^2 
\left ( s_0 - \frac{\gm_1\zeta}{\gm_1 + \gm_3} 
\right ) \left ( 2s_0 - \frac{\gm_1\zeta}{\gm_1 + \gm_3} 
\right ) t^3 \;  .
\ee
>From Eq. (79), it follows that the population difference
does not essentially vary during this stage, being close
to $s_0$. Taking into account that $\gamma_1 \ll \gamma_3$,
we find that the coherence function (80) is either linear
in time or cubic in time,
$$
w \simeq 2\gm_3 s_0^2 t \qquad (s_0\neq 0)  \; , 
$$
\be
\label{81}
w \simeq \frac{2}{3}\gm_1^2 \gm_3 \zeta^2 t^3 \qquad (s_0=0)\; ,
\ee
depending on whether there exists or not the initial
polarization $s_0 \equiv s(0)$. The value of the function
$w$ in the former case, if $s_0 \sim 1$, is much larger than
the value of $w$ in the second case. In the later case, the
value of $w$ is practically negligible. This shows that in
order that coherence could really develop, it is necessary
to have sufficient initial population difference $s_0$.

During the quantum stage, the evolution is mainly due to
random quantum fluctuations corresponding to the term
$\gamma_3 s^2$, while coherence starts being noticeable,
when the term $\gamma_2(gs - 1)w$, responsible for
collective effects, becomes of the same order as the
quantum term. That is, the coherence time $t_{coh}$ can be
estimated from the equality
\be
\label{82}
\gm_2 ( gs - 1 ) w = \gm_3 s^2 \qquad (t = t_{coh})  \; ,
\ee
when the quantum and collective terms coincide. This equality
can hold only when $gs > 1$. Since $s$ does not vary much
during the quantum stage, the condition for the existence of
the coherence time can be written as
\be
\label{83}
gs_0 > 1 \;  ,
\ee
which implies that $s_0$ must be positive. Equation (82)
gives the coherence time
\be
\label{84}
t_{coh} = 
\frac{s_0/2}{\gm_2(gs_0-1)s_0 + \gm_3s_0+ \gm_1(s_0-\zeta)} \; .
\ee
In the standard situation, when $\gamma_1 \ll \gamma_2$ and
$\gamma_2 \sim \gamma_3$, the coherence time (84)
reduces to
\be
\label{85}
t_{coh} = \frac{1/2}{\gm_2(gs_0-1)+\gm_3} \; .
\ee
For sufficiently strong coupling, the coherence time is
\be
\label{86}
t_{coh} \simeq \frac{T_2}{2gs_0} \qquad (gs_0 \gg 1) \; .
\ee
To estimate the coherence time, let us take $gs_0 \sim 10$.
With the dephasing time $T_2 \sim 10^{-13}s - 10^{-12}s$, we
get $t_{coh} \sim 10^{-14}s - 10^{-13}s$. At the end of the
quantum stage, solutions (79) and (80) are well approximated
by the forms
\be
\label{87}
w(t_{coh}) \simeq 2\gm_3 t_{coh} s_0^2 \;, \qquad
s(t_{coh}) \simeq s_0 \; .
\ee
The coherence function $w$ here is yet very small, being of
order $1/ gs_0 \ll 1$, and the population difference is yet
close to the initial value $s_0$.

\vskip 5mm
{\bf C. Coherent Stage}

\vskip 2mm
After the coherence time, collective effects become dominant.
Coherence can last during the time interval
\be
\label{88}
t_{coh} < t < T_2 \; .
\ee
At this stage, taking into account that $T_2 \ll T_1$, hence,
$\gamma_1 \ll \gamma_2$, Eqs. (70) take the form
\be
\label{89}
\frac{dw}{dt} = -2\gm_2 ( 1-gs) w \; , \qquad
\frac{ds}{dt} = -g\gm_2 w \; .
\ee
These equations enjoy the exact solutions describing the
superradiant pulse
\be
\label{90}
w = \left ( \frac{\gm_p}{g\gm_2} \right )^2 {\rm sech}^2
\left ( \frac{t-t_0}{\tau_p} \right ) \;  , \qquad
s = \frac{1}{g} \; - \; \frac{\gm_p}{g\gm_2} \tanh 
\left ( \frac{t-t_0}{\tau_p} \right ) \; ,
\ee
where the integration constants $\tau_p \equiv 1/\gamma_p$
and $t_0$ are defined by the initial conditions (87). The
{\it pulse width} $\gamma_p$ is given by the relations
\be
\label{91}
\gm_p^2 = \gm_g^2 + 2(g\gm_2)^2\gm_3 t_{coh} s_0^2 \;  ,
\qquad \gm_g \equiv (gs_0-1) \gm_2 \; ,
\ee
which, keeping in mind that $\gamma_3 t_{coh} \ll 1$, yields
the {\it pulse time}
\be
\label{92}
\tau_p = \frac{T_2}{gs_0-1} \left [ 1 \; - \;
\frac{\gm_3 t_{coh}g^2 s_0^2}{(gs_0-1)^2} \right ] \; .
\ee
The second integration constant is the {\it delay time}
\be
\label{93}
t_0 = t_{coh} + \frac{\tau_p}{2} \; 
\ln \left | \frac{\gm_p+\gm_g}{\gm_p -\gm_g} \right | \; ,
\ee
corresponding to the maximum of the pulse. In view of the
inequality $\gamma_3 t_{coh} \ll 1$, the pulse width can be
represented as
\be
\label{94}
\gm_p = (gs_0-1)\gm_2 + 
\frac{g^2\gm_2\gm_3 t_{coh} s_0^2}{gs_0-1} \;  ,
\ee
where condition (83) is taken into account. Then the delay
time (93) is
\be
\label{95}
t_0 = t_{coh} + \frac{\tau_p}{2} \; \ln \left |
\frac{2(gs_0-1)^2}{g^2\gm_3t_{coh}s_0^2} \right | \; .
\ee
In the case of strong coupling $gs_0 \gg 1$, we have
\be
\label{96}
t_0 \simeq t_{coh} + t_{coh} \ln \left |
\frac{2}{\gm_3 t_{coh}} \right | \;  ,
\ee
with $\tau_p \simeq 2t_{coh}$, and the coherence time is given 
by Eq. (86). For $\gamma_3\sim\gamma_2\sim 10^{12} Hz-10^{13}Hz$
and $t_{coh} \sim 10^{-14}s - 10^{-13}$s, we have
$\gamma_3 t_{coh} \sim 0.01 - 0.1$. Then $t_0 \sim 5t_{coh}$.
For the coupling $gs_0 \gg 1$, the pulse width, as follows from
Eq. (92), is $\tau \simeq T_2/gs_0$, which is inversely
proportional to the dot density $\rho$, that is, it is inversely
proportional to the number of dots taking part in the radiation
process. This is a typical feature of superradiant emission.

The described coherent radiation arises as a self-organized 
process caused by the dot interactions through the common 
radiation field. At the initial time no coherence has been 
imposed on the system, so that $w(0)=0$. The radiation process 
is triggered by the transition dipolar waves, and coherence 
develops from the initially incoherent chaotic stages. This 
process of coherence, self-consistently arising from chaos, is 
the most interesting and the most difficult for description. 
The appearing superradiant emission is called {\it pure
superradiance}.

The situation is much simpler, when coherence is imposed on 
the system from the very beginning, by means of an external 
field, such that $w(0)\neq 0$. If the system is coherent 
starting from $t=0$, the incoherent stages do not exist, which 
implies that $t_{coh}$ is zero. The resulting coherent emission 
corresponds to the {\it triggered superradiance}. The superradiant
pulse is described by the solutions of the same form (90), but 
with the pulse width given by the expression
$$
\gm_p^2 = \gm_g^2 + (g\gm_2)^2 w_0 \; ,
$$
where $w_0\equiv w(0)\neq 0$. Then the pulse time is
$$
\tau_p = 
\frac{1}{\sqrt{\gm_g^2 + (g\gm_2)^2 w_0} } \; .
$$
For sufficiently strong coupling, such that $gs_0\gg 1$, the 
pulse time becomes
$$
\tau_p \simeq \frac{T_2}{g\sqrt{s_0^2+w_0} } \; .
$$
In the case of the triggered superradiance, the pulse time 
depends both on the initial population inversion as well as 
on the level of the imposed coherence.

\vskip 5mm

{\bf D. Relaxation Stage}

\vskip 2mm
In the time interval
\be
\label{97}
T_2 < t \ll T_1 \; ,
\ee
when also $t \gg t_0$, the coherent solutions (90) decay as
\be
\label{98}
w \simeq \left ( \frac{2\gm_p}{g\gm_2}\right )^2
\exp \left ( - \; \frac{2t}{\tau_p}\right ) \; , \qquad
s \simeq \frac{\gm_2-\gm_p}{g\gm_2} + \frac{2\gm_p}{g\gm_2}\; 
\exp \left ( -\; \frac{2t}{\tau_p}\right )\; .
\ee
Coherence dies out and the population difference relaxes to
\be
\label{99}
s  \simeq \frac{\gm_2-\gm_p}{g\gm_2} \qquad 
( t \gg t_0 ) \;  ,
\ee
corresponding to $s$ inverted as compared to its initial
value $s_0$. For strong coupling $gs_0 \gg 1$, when
$\tau_p \simeq 2 t_{coh}$, $t_{coh} \simeq 1/(2\gamma_2 gs_0)$,
and $\gamma_p \simeq \gamma_2 g s_0$, expression (99)
equals $-s_0$, which implies practically complete inversion.

\vskip 5mm
{\bf E. Stationary Stage}

\vskip 2mm
In the present subsection, we assume that either there is no
permanent pumping, so that $\zeta = -1$ or that the external
pumping is weak, such that $|g \zeta| \ll 1$. Then, at
asymptotically large time
\be
\label{100}
t \gtrsim T_1 \; ,
\ee
the system tends to its stationary state. All terms of Eq. (70)
play role at this stage. That is, the evolution is described by
the equations
\be
\label{101}
\frac{dw}{dt} = - 2\gm_2(1-gs) w + 2\gm_3s^2 \; , \qquad
\frac{ds}{dt} = -g\gm_2 w -\gm_3 s - \gm_1(s-\zeta) \; .
\ee
In order to find out the stable stationary solutions, it is
necessary to resort to the Lyapunov stability analysis. For
this purpose, we calculate the Jacobian matrix
$\hat{J}(t) = [J_{ij}(t)]$, whose elements are
$$
J_{11} \equiv \frac{\prt}{\prt w}\left ( \frac{dw}{dt}
\right ) =  2\gm_2(gs-1)  \; ,  \qquad
J_{12} \equiv \frac{\prt}{\prt s}\left ( \frac{dw}{dt}
\right ) =  2\gm_2gw + 4\gm_3 s  \; ,  
$$
$$
J_{21} \equiv \frac{\prt}{\prt w}\left ( \frac{ds}{dt}
\right ) =  - g\gm_2  \; ,  \qquad
J_{22} \equiv \frac{\prt}{\prt s}\left ( \frac{ds}{dt}
\right ) =  -\gm_1 - \gm_2  \; .
$$
The stationary solutions are given by the zeros of the
right-hand sides of Eqs. (101). Evaluating the Jacobian matrix
at the fixed points, we analyze the stability of the latter.
Below, only the stable stationary solutions are presented.

In the case when $g\zeta \ll -1$, the stable stationary
solutions are
\be
\label{102}
w^* \simeq \frac{\gm_3|\zeta|}{\gm_2|g|} \; , \qquad
s^* \simeq \zeta \left ( 1 - \;
\frac{\gm_3}{\gm_1|g\zeta|} \right ) \;  ,
\ee
which correspond to a stable node, since the eigenvalues of
the Jacobian matrix, defining the characteristic exponents,
are all negative,
$$
 J_1 \simeq -\gm_1 - \gm_3 \; , \qquad
J_2 \simeq -2\gm_2|g\zeta| \; .
$$
The coherence function $w$ is very small.

For weak pumping, such that $|g\zeta| \ll 1$, the stable
stationary solutions are
$$
w^* \simeq \left ( \frac{\gm_1\zeta}{\gm_1+\gm_3}\right )^2 \;
\frac{\gm_3}{\gm_1} \; \left [ 1 +
\frac{\gm_1(\gm_1-\gm_3)g\zeta}{(\gm_1+\gm_3)^2} \right ] \;  ,
$$
\be
\label{103}
s^* \simeq \frac{\gm_1\zeta}{\gm_1+\gm_3} \; 
\left [ 1 -\;
\frac{\gm_1\gm_3g\zeta}{(\gm_1+\gm_3)^2} \right ] \;  ,
\ee
which also correspond to a stable node, as far as the
characteristic exponents are
$$
J_1 \simeq -\gm_1 - \gm_3 \; , \qquad
J_2 \simeq - 2\gm_2 \; .
$$
Because of the relations $\gamma_1 \ll \gamma_2 \sim \gamma_3$,
the level of coherence is very small, that is, $w^* \ll 1$.
At the stationary stage, when there is no external pumping,
coherence is practically absent, since it has been died out
yet during the relaxation stage. The population difference at
this stage is close to $\gamma_1 \zeta / \gamma_3$.

\vskip 5mm
{\bf F. Pulsing Superradiance}

\vskip 2mm
In the case when there is a sufficiently strong external pumping,
such that $g\zeta \gg 1$, the fixed points
\be
\label{104}
w^* \simeq \frac{\gm_1\zeta}{\gm_2g} \; , \qquad 
s^* \simeq \frac{1}{g} \left ( 1 - \; 
\frac{\gm_3}{\gm_1 g\zeta}\right ) \; ,
\ee
represent a stable focus, with the characteristic exponents
$$
 J_{1,2} \simeq -\; \frac{1}{2}\; 
(\gm_1 + \gm_3) \pm i\om_{eff} \; ,
$$
in which the effective asymptotic frequency is
$$
 \om_{eff} \equiv \; \sqrt{2g\zeta\gm_1\gm_2} \; .
$$
The effective asymptotic frequency $\omega_{eff}$ defines 
the {\it effective asymptotic period}
\be
\label{105}
T_{eff} \equiv \frac{2\pi}{\om_{eff}} = 
\pi\; \sqrt{ \frac{2T_1 T_2}{g\zeta} }	\; .
\ee
In this regime, there occurs a series of superradiant pulses,
bursting in the intervals of time close to the effective
period (105). The total number of such pulses is of order
$T_1 / T_{eff}$. In the presence of a permanent nonresonant 
pumping, guaranteeing the value $\zeta \sim 1$, the quantity 
$\gamma_1$ acquires the meaning of the pumping rate, which can 
be made comparable with $\gamma_2$. Therefore, the number of 
pulses is of order $\sqrt {g \zeta}$. Hence, there can be 
produced several, around 10, pulses. The interval between the 
superradiant pulses is of order $T_{eff} \sim 10^{-13}s$.

\vskip 5mm
{\bf G. Numerical Illustration}

\vskip 5mm
In order to illustrate the dynamics of radiation in graphical 
form, we calculate numerically the quantities $w$ and $s$ as 
functions of time for some parameters typical of quantum dots.
To evaluate these parameters, we use the values from Sec. II, 
from where we have $\gamma_0/\gamma_2 \sim 10^{-2}-10^{-3}$ 
and $\gamma_1/\gamma_2\sim 10^{-4}-10^{-3}$. The value of the 
coupling parameter $g$, given in Eq. (73), is 
$g \sim (\gamma_0/\gamma_2)\rho\lambda^3$. The coherence factor
$\rho\lambda^3 \sim 10-10^5$. Thence, $g\sim 10^{-2}-10^3$. 
According to condition (83), well developed coherence appears
when $gs_0 > 1$. Since the initial condition $s_0$ cannot be 
larger than 1, it should be that $g>1$. And, as is mentioned 
above, for semiconductors, $\gamma_3 \geq \gamma_2$.

Figures 1, 2, and 3 show the evolution of $w=w(t)$ and $s=s(t)$ 
as functions of dimensionless time $t$, measured in units of 
$T_2 \equiv 1/ \gamma_2$, for several typical cases. We assume 
that at the initial time, the system is inverted, but coherence 
is absent and develops in a self-organized way. Recall that the 
standard mean-field, or semiclassical, approximation cannot 
describe such a self-organized appearance of coherence. Figures
1 and 2 correspond to the case of no external pumping, when 
$\zeta = -1$. The difference between these figures is in the 
value of the dynamical attenuation rate. As we see, the larger
$\gamma_3$ decreases the delay time and makes the superradiant 
pulse strongly asymmetric. Recall that for atoms in free space,
$\gamma_3$ is usually much smaller than $\gamma_2$, because of 
which superradiant pulses produced by atoms are more symmetric.
The essential asymmetry of superradiant pulses is the feature
typical of quantum-dot radiation. Another typical feature of 
the quantum-dot dynamics, also caused by the large rate 
$\gamma_3$, is the much faster, than for atoms in free space,
tendency of the population difference to the stationary state.  
Figure 3 demonstrates the radiation dynamics in the case of an
external pumping, when $\zeta = 1$, and there appear several 
superradiant pulses with decaying amplitude.

\section{Conclusion}

The theory of quantum-dot radiation is developed being based
on microscopic equations. The possibility of realizing the
superradiant regime is analyzed. The temporal evolution during
all radiation stages is studied in detail. A special attention
is payed to the process when coherence arises from an initially
incoherent state. The description of this process is impossible
by means of the standard semiclassical equations, because of
which a more accurate method has been used in the paper,
employing the stochastic mean-field approximation that has
been developed earlier and applied for describing the dynamics 
of spin assemblies [5-11], Bose systems in random fields [38-40], 
and atomic squeezing [50].

It is necessary to emphasize that the radiation dynamics of 
quantum dots has several specific features distinguishing this
dynamics from atomic radiation. This is connected, first of all, 
with rather different values of physical dot parameters, as 
compared to atomic parameters. Because of this, despite many 
analogies, the theory of dot radiation requires a separate 
investigation. The principal theoretical points that have been 
suggested in the present paper for the adequate description of
dot radiation are as follows.
   
(i) Because of essential current fluctuations in semiconductor, 
the standard semiclassical approximation, often used for atoms in 
free space, is not applicable for quantum dots. For the latter more 
elaborate techniques are required, such as the stochastic mean-field 
approximation.

(ii) Taking into account the fluctuation of current makes the 
dynamic attenuation parameter $\gamma_3$ of the order or 
larger than $\gamma_2$. This is contrary to the case of atoms in 
free space, where usually $\gamma_2$ is the largest relaxation 
parameter. 

(iii) For the correct description and principal understanding of 
the mechanism, triggering the beginning of the radiation process,
it is important to stress the existence of triggering dipolar waves.

(iv) The single-mode picture is not applicable for quantum dots. 
It is necessary to consider a bunch of transverse modes forming 
spatial filaments. To reduce the consideration to a treatable 
problem, it is necessary to involve some tricks, like the 
transverse-mode expansion.

(v) The overall dynamics of dot radiation consists of several 
stages, which have been thoroughly studied and described, both
analytically and numerically, for the parameters typical of 
quantum dots.  

In the dynamics of dot radiation, it is possible to distinguish 
the following qualitatively different stages. The first is the
{\it fluctuating stage} lasting during the time interval
$0 < t < t_{int}$, when the radiation process is triggered by
dipolar waves. At this stage, there is no yet sufficiently
strong interaction between dots. The interaction time is of
order $t_{int} \sim 10^{-15} s - 10^{-14} s$.

The second is the {\it quantum stage} in the temporal interval
$t_{int} < t < t_{coh}$, when the dot interactions through
photon exchange start playing noticeable role, but coherence
has not yet been developed. The coherence time, required for
the appearance of well developed coherence, is of order
$t_{coh} \sim 10^{-14} s - 10^{-13} s$.

Then the {\it coherent stage} comes into play in the interval
$t_{coh} < t < T_2$, when the dots emit a coherent
superradiant pulse. For the quantum dot materials, the
dephasing time is of order $T_2 \sim 10^{-13} s - 10^{-12} s$.
The maximum of the pulse occurs at the delay time
$t_0 \simeq 5 t_{coh}$ and the pulse duration is
$\tau_p \simeq 2 t_{coh}$. The pulse duration is inversely
proportional to the dot density, that is, inversely
proportional to the number of dots, involved in the process
of radiation, which is a typical feature of superradiance.

After the superradiant pulse is emitted, the system relaxes
to an incoherent state during the {\it relaxation stage} in
the interval $T_2 < t \ll T_1$. The population difference
reverses. For the system of dots in a semiconducting material, 
the longitudinal relaxation time is $T_1 \sim 10^{-9} s$. 
But this is not yet the final stage of evolution.

The {\it stationary stage} is reached for $t \gtrsim T_1$,
if there is no external permanent pumping or the effective
dot interactions are weak, so that $|g\zeta|\ll 1$. Then the
system tends to a stationary incoherent state representing a
stable node.

If the system of dots is subject to a sufficiently strong
external permanent pumping, such that $|g\zeta| \gg 1$, the
regime of {\it pulsing superradiance} occurs. Then a series
of about 10 superradiant bursts can appear, flashing in the 
intervals of time $T_{eff} \sim 10^{-13}s$.

\vskip 5mm
{\bf Acknowledgement}

\vskip 3mm
Financial support from the Russian Foundation for Basic
Research is appreciated.

\vskip 5mm
{\bf Appendix}

\vskip 2mm

Here, the explanation is given of the contributions coming 
from the dot self-action. For a dot located at $\bf r = 0$, 
the self-action vector potential is
$$
A_{self}^\alpha(\br,t) = \frac{1}{c} \sum_\beta \int
\frac{\dlt_{\al\bt}(\br')}{|\br-\br'|} \;
J^\beta \left ( t - \; \frac{|\br-\br'|}{c} \right ) \; d\br' \; , 
\hspace{4cm} (A1)
$$
with the current
$$
\bJ \left ( t - \; \frac{r}{c} \right ) =
i\om_0 \left [ \bd S^+(t) e^{-ik_0 r} -
\bd^* S^-(t) e^{ik_0r} \right ] \Theta(ct-r) \;  ,  
\qquad\quad \qquad \qquad \qquad (A2)
$$
in which $S^\alpha(t) \equiv S^\alpha(0,t)$. At short distance,
such that $k_0 r \ll 1$, one has $\exp(ik_0 r) \simeq 1 + ik_0 r$.
Substituting the transverse delta-function (9) into the vector
potential (A1), we keep in mind that the averaging of the dipolar
tensor over spherical angles yields zero. As a result, the vector
potential (A1) becomes
$$
\bA_{self}(\br,t) = \frac{2}{3}\; k_0^2 \left [ \bd S^+(t) +
\bd^* S^-(t) \right ] + i\; \frac{2k_0}{3r}
\left [ \bd S^+(t) - \bd^* S^-(t) \right ]\;  .
\hspace{2cm} (A3)
$$
Averaging this expression over the radial variable between the
electron wavelength $\lambda_e = 2\pi/mc$, with $m$ being the
electron mass, and the radiation wavelength $\lambda = 2\pi/k_0$,
and taking into account that $\lambda_e \ll \lambda$, we get the
self-action potential
$$
\bA_{self}(t) = \frac{2}{3}\; k_0^2 \left [ \bd S^+(t) +
\bd^* S^-(t) \right ] + \frac{ik_0}{3\pi}\; \ln \left (
\frac{mc^2}{\om_0} \right ) \left [ \bd S^+(t) -
\bd^* S^-(t) \right ]   .
\hspace{1cm} (A4)
$$

Let us introduce the natural width
$$
\gm_0 \equiv \frac{2}{3}\; |\bd|^2 k_0^2
\hspace{12cm} (A5)
$$
and the Lamb shift
$$
\dlt_L \equiv \frac{\gm_0}{2\pi} \;
\ln \left ( \frac{mc^2}{\om_0} \right )\;  .
\hspace{10.5cm} (A6)
$$
Then the terms in Eqs. (14), induced by the self-action potential
(A4), are
$$
2k_0 S^z \bd \cdot \bA_{self} = ( i \dlt_L - \gm_2) S^- +
\frac{\bd^2}{|\bd|^2} \; ( \gm_2 + i\dlt_L ) S^+
\hspace{5.2cm} (A7)
$$
for the first of Eqs. (14) and
$$
k_0 \left ( S_i^+ \bd + S_i^- \bd^* \right ) \cdot
\bA_{self} = \gm_1 \left ( \frac{1}{2} + S^z \right )
\hspace{7.1cm} (A8)
$$
for the second, where $\gamma_1 = 2 \gamma_0$ and
$\gamma_2 = \gamma_0$. In the standard situation, one has
$$
 \frac{\gm_0}{\om_0} \ll 1 \; , \qquad
\frac{\dlt_L}{\om_0} \ll 1\;  .
\hspace{10.2cm} (A9)
$$
The Lamd shift, without the loss of generality, can be included
in the definition of the transition frequency $\omega_0$.

\newpage

\begin{center}
{\Large {\bf Figure Captions}}

\end{center}

\vskip 3cm

Fig. 1. The coherence intensity $w$ (solid line) and population 
difference $s$ (dashed line) as functions of dimensionless time 
(measured in units of $T_2$) for the attenuation parameters
$\gamma_1 = 0.003, \gamma_3 = 1$ (measured in units of $\gamma_2)$,
for the coupling parameter $g = 10$, with the initial conditions
$w_0 =0, s_0 = 1$.  

\vskip 3cm

Fig. 2. The coherence intensity $w$ (solid line) and population 
difference $s$ (dashed line) as functions of dimensionless time 
(measured in units of $T_2$) for the attenuation parameters
$\gamma_1 = 0.003, \gamma_3 = 10$ (measured in units of $\gamma_2)$,
for the coupling parameter $g = 10$, with the initial conditions
$w_0 =0, s_0 = 1$. The larger dynamic attenuation $\gamma_3$ makes
the pulse more asymmetric.

\vskip 3cm

Fig. 3. The coherence intensity $w$ (solid line) and population 
difference $s$ (dashed line) as functions of dimensionless time 
(measured in units of $T_2$) in the case of an external pumping, for 
the parameters $\gamma_1 = 10, \gamma_3 = 1$ (measured in units of 
$\gamma_2)$, for the coupling parameter $g = 100$, with the initial 
conditions $w_0 =0, s_0 = 1$. The coherence intensity, as well as
population difference, exhibit five pulses with decaying amplitude.

\newpage

\begin{figure}[h]
\centerline{\includegraphics[width=16cm]{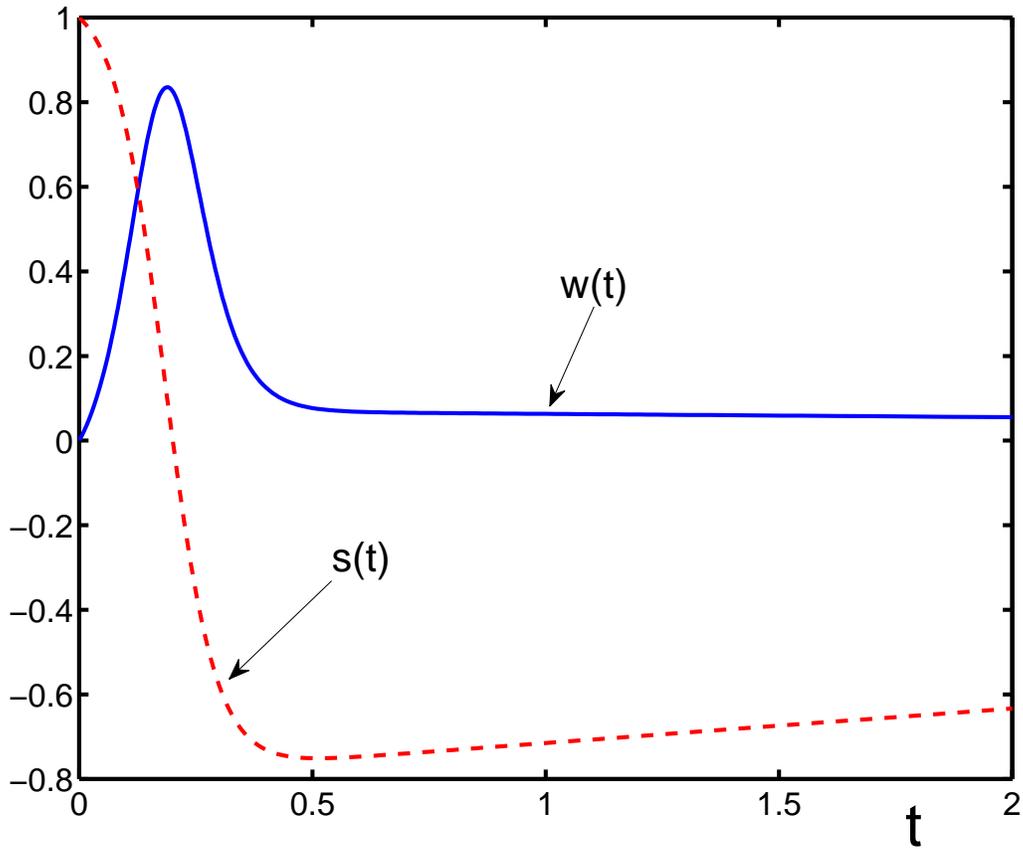}}
\caption{The coherence intensity $w$ (solid line) and population 
difference $s$ (dashed line) as functions of dimensionless time 
(measured in units of $T_2$) for the attenuation parameters
$\gamma_1 = 0.003, \gamma_3 = 1$ (measured in units of $\gamma_2)$,
for the coupling parameter $g = 10$, with the initial conditions
$w_0 =0, s_0 = 1$.}
\label{fig:Fig.1}
\end{figure}

\newpage

\begin{figure}[h]
\centerline{\includegraphics[width=16cm]{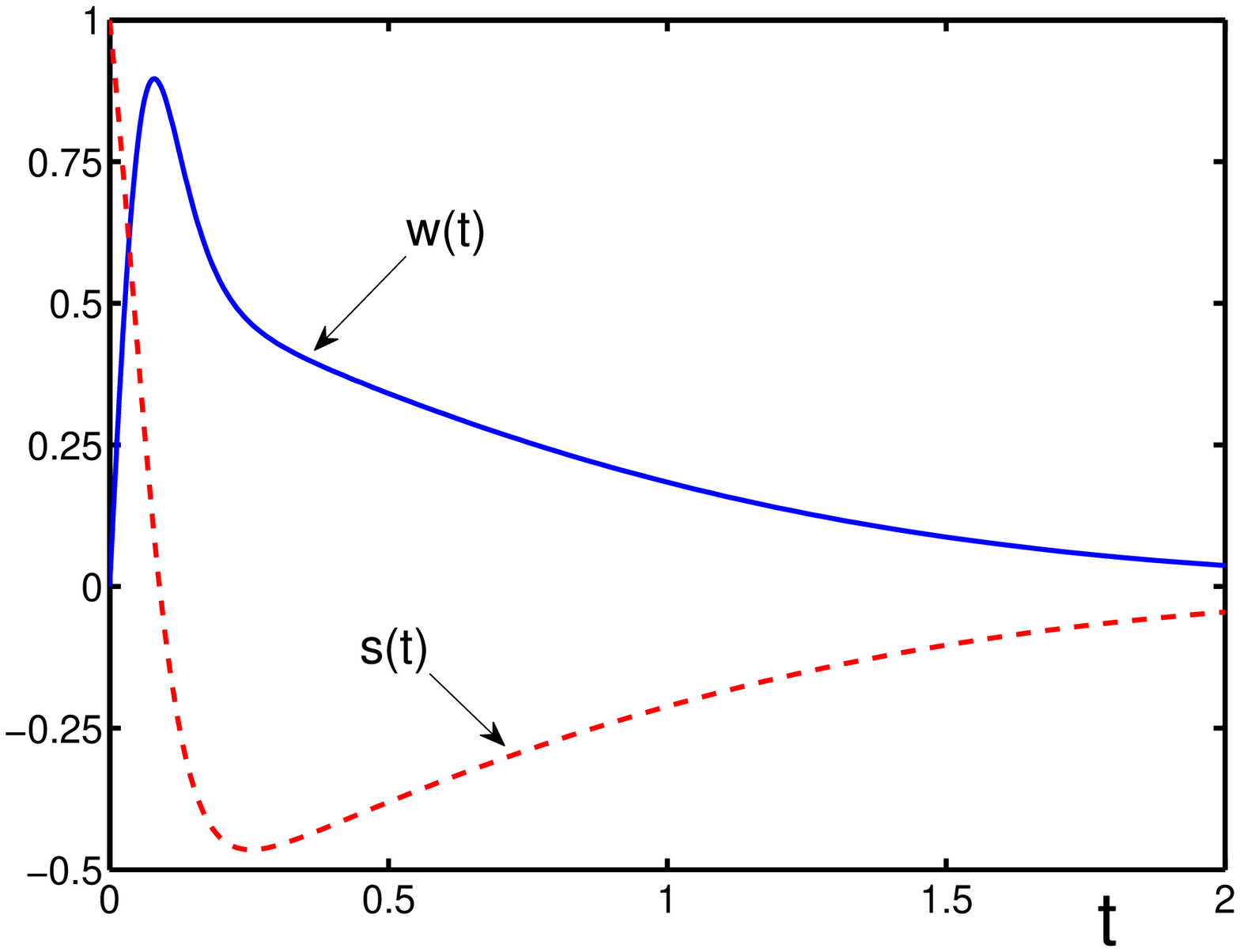}}
\caption{The coherence intensity $w$ (solid line) and population 
difference $s$ (dashed line) as functions of dimensionless time 
(measured in units of $T_2$) for the attenuation parameters
$\gamma_1 = 0.003, \gamma_3 = 10$ (measured in units of $\gamma_2)$,
for the coupling parameter $g = 10$, with the initial conditions
$w_0 =0, s_0 = 1$. The larger dynamic attenuation $\gamma_3$ makes
the pulse more asymmetric.}
\label{fig:Fig.2}
\end{figure}

\newpage

\begin{figure}[h]
\centerline{\includegraphics[width=16cm]{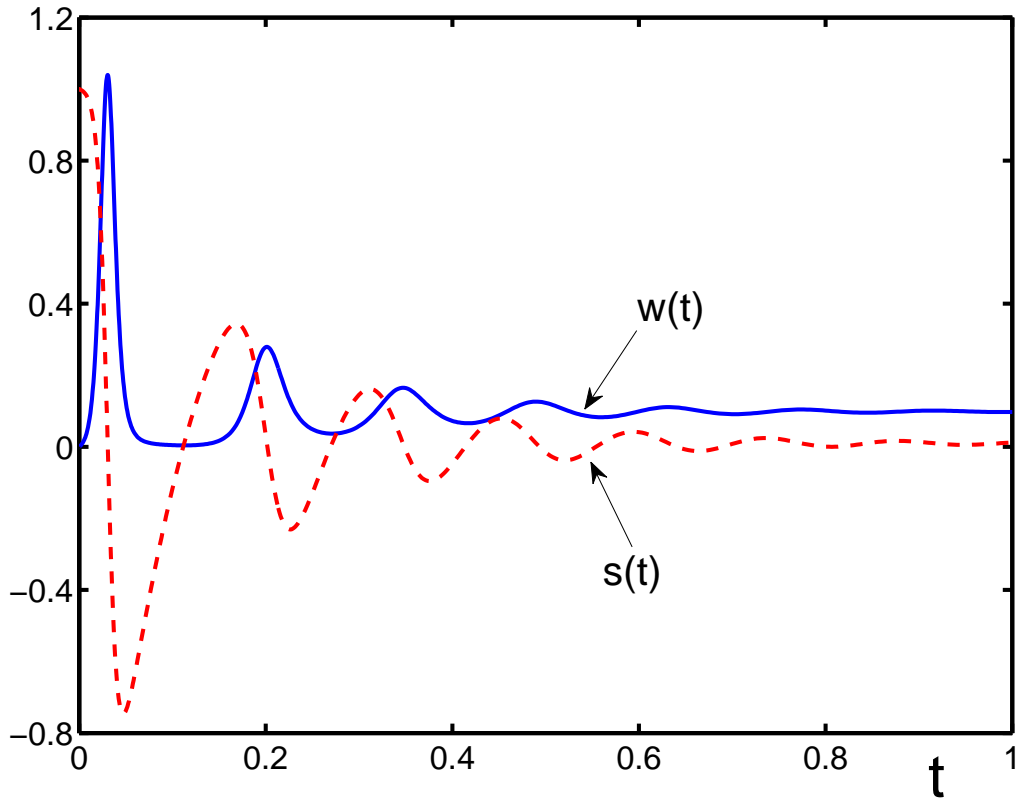}}
\caption{The coherence intensity $w$ (solid line) and population 
difference $s$ (dashed line) as functions of dimensionless time 
(measured in units of $T_2$) in the case of an external pumping, for 
the parameters $\gamma_1 = 10, \gamma_3 = 1$ (measured in units of 
$\gamma_2)$, for the coupling parameter $g = 100$, with the initial 
conditions $w_0 =0, s_0 = 1$. The coherence intensity, as well as
population difference, exhibit five pulses with decaying amplitude.}
\label{fig:Fig.3}
\end{figure}

\end{document}